# Oscillatory finite-time singularities in rockbursts

Qinghua Lei[1,*], Didier Sornette[2]

[1]*Department of Earth Sciences, Uppsala University, Uppsala, Sweden*

[2]*Institute of Risk Analysis, Prediction and Management, Academy for Advanced Interdisciplinary Studies, Southern University of Science and Technology, Shenzhen, China*

**Abstract**

Forecasting violent rockbursts remains a formidable challenge due to significant uncertainties involved. One major uncertainty arises from the intermittency of rock failure processes, typically characterised by a series of progressively shorter quiescent phases punctuated by sudden accelerations, rather than a smooth continuous progression towards the final breakdown. This non-monotonic evolution of rock mass deformation complicates rockburst prediction, challenging conventional time-to-failure models that often assume a smooth power law accelerating behaviour. Here, we introduce a generalised time-to-failure model called log-periodic power law singularity (LPPLS) model to effectively capture the intermittent dynamics of damage and rupture processes in rock leading up to violent rockbursts. We perform parametric and nonparametric tests on 11 historical rockburst events at three underground mines, documenting empirical evidence and providing theoretical arguments to demonstrate the significance of log-periodic oscillatory power law finite-time singularities. Log-periodicity in these rockburst events is likely driven by the interaction of subparallel propagating cracks, the diffusion of stress-triggering processes, or the interplay between stress drop and stress corrosion. Our results and insights obtained have significant implications for not only understanding but also forecasting rockbursts, as recognising and characterising log-periodicity can help transform intermittency from traditionally perceived noise into valuable predictive information.

* Corresponding author: qinghua.lei@geo.uu.se

1. Introduction

Rockbursts have long been a major challenge in rock mechanics and mining sciences[1–5] and their importance continues to grow as the extraction of Earth's resources (e.g. metals, rare earth elements, coal) as well as the construction of infrastructures (e.g. tunnels and caverns) extends to ever greater depths[6]. Rockbursts can be in general classified into two major types[7,8]: (i) strainbursts, which result from the accumulation and sudden release of strain energy in local rock under high stress concentrations, and (ii) fault-slip bursts, which are caused by shear slip along remote faults. Reliable rockburst forecast is essential for mining or construction companies to inform underground workers about potential rockburst events and their timing, enabling timely evacuations and the implementation of necessary safety measures. Great efforts have been dedicated to developing and deploying high-precision monitoring technologies to record rock mass deformations and associated seismic activities during deep underground excavations[9,10], aiming to detect and characterise potential precursors of impending rockburst events. However, the current practice of rockburst hazard analysis largely relies on empirical criteria[11]. For example, rockburst liability is often defined based on the ratio of stress level to rock strength against empirically defined thresholds[12]; many early warning systems are based on in-situ microseismic monitoring and empirically obtained relations between microseismicity parameters and rockburst occurrences[13]. However, the underlying physical mechanisms behind these empirical relationships usually remain elusive such that their application is often plagued by large uncertainties. Moreover, these approaches are akin to drawing conclusions from static snapshots, whereas a more effective strategy would be a movie-like approach, capturing the full temporal evolution of the process leading to rupture. On the other hand, extensive studies have been conducted to understand the physical processes driving rockbursts based on high-fidelity numerical simulations[14–19], but the complexity of computational modelling and the difficulty in model calibration against observational data hinder their practical application for rockburst forecasting.

Over the past decades, significant advances have been made to forecast catastrophic failures with more physically based models proposed and tested. For example, the power law time-to-failure model, which was originally developed based on empirical observations[20–22] but later backed by physical arguments[23–28], has been widely adopted to predict the timing of various catastrophic events such as landslides[29] and volcanic eruptions[30] as well as rockbursts[31]. This model exemplifies the movie-like approach by incorporating the complete acceleration history leading up to rupture. Several physical mechanisms have been proposed to explain this commonly-observed power law time-to-failure dynamics. One is based on the analogy drawn between catastrophic material failures and critical points in thermodynamics, heuristically linking the emergence of power laws during material rupture to those near critical phase transitions[32,33]. Other possible mechanisms include stress corrosion damage[23–26] and frictional sliding instability[27,28]. However, large uncertainties have been found in applying this power law model for time-to-failure prediction[34,35], primarily due to the sporadic nature of rupture phenomena, which challenges the underlying assumption of a smooth, monotonic power law acceleration behaviour.

The Log-Periodic Power Law Singularity (LPPLS) model extends the conventional power law framework by incorporating log-periodic components, enabling it to capture the intermittent rupture dynamics in heterogeneous systems[36–38], as demonstrated for earthquakes[39], rockbursts[40], and glacier breakoffs[41]. By extensively testing on a global dataset of various geohazard events including landslides, rockbursts, glacier breakoffs, and volcanic eruptions, we have recently proven that the LPPLS model significantly outperforms the conventional power law model[42]. This is because the LPPLS model can leverage the unsteady, non-monotonic signals inherent in rupture dynamics to refine its description and prediction of material failure. These signals of intermittency—traditionally perceived as noise—intrinsically arise from the localised and threshold nature of the mechanics of rupture in heterogeneous materials[38,43].

This technical note aims to present further empirical evidence regarding the significance of log-periodicity in rockbursts and develop theoretical insights into the underlying mechanisms that drive rockbursts. The remainder of this note is organised as follows. In section 2, we introduce the mathematical formulation of the LPPLS model together with the procedures for parametric calibration and nonparametric examination. In section 3, we perform both parametric and nonparametric tests on 11 historical rockburst events documented at 3 underground mines. Finally, in section 4, we assess the statistical significance of log-periodicity in rockbursts, explore the underlying physical mechanisms, and discuss the implications for rockburst prediction.

## 2. Methodology

The rate-dependent rock failure behaviour often obeys the following nonlinear relation[21,22]:

$$\frac{d^2\Omega}{dt^2} = \mu\left(\frac{d\Omega}{dt}\right)^\alpha, \text{ with } \alpha > 1, \qquad (1)$$

where $\Omega$ is an observable quantity (e.g. displacement, strain, and energy release), $t$ is time, $\mu > 0$ is a constant, and $\alpha$ is an exponent measuring the degree of nonlinearity. The condition of $\alpha > 1$ guarantees the appearance of a finite-time singularity, which becomes evident by integrating equation (1) to have:

$$\frac{d\Omega}{dt} = \frac{\gamma}{(t_c - t)^\beta}, \qquad (2)$$

where $\gamma = (\beta/\mu)^\beta$, $\beta = 1/(\alpha-1)$, with $\beta > 0$ (for $\alpha > 1$) ensuring a singular behaviour as the critical time $t_c$ is approached. A further integration of equation (2) yields the so-called power law time-to-failure model[23,44]:

$$\Omega(t) = A + B(t_c - t)^m, \text{ with } m < 1, \qquad (3)$$

where $m = 1-\beta = (\alpha-2)/(\alpha-1)$ is the critical exponent, and $A$ and $B = -\gamma/m$ are constants. This power law relationship shows continuous scale invariance, meaning that scaling $t_c-t$ by an arbitrary factor $\lambda$ causes the difference between the observable and the value $A$ to scale by a factor of $\lambda^m$, independent of $t_c-t$.

Let us now consider a generalised scenario, where the critical exponent is extended from real to complex numbers, written as $m+i\omega$. This generalisation, anticipated in dissipative systems with out-of-equilibrium dynamics and quenched disorder[45], is evident from the general solution of $\Omega$ to renormalization group equations near critical points[39,46]. Keeping the first Fourier expansion of this general solution leads to the following LPPLS formula[36,37]:

$$\Omega(t) = A + \{B + C\cos[\omega\ln(t_c - t) - \phi]\}(t_c - t)^m, \text{ with } m < 1. \qquad (4)$$

which introduces a log-periodic component associated with a relative amplitude of $C/B$, an angular log frequency $\omega$, and a phase shift $\phi$, oscillating around the overall power law characterised by the amplitude factor $B$. The local maxima of this log-periodic component form a geometric time series $\{t_1, t_2, \ldots, t_k, \ldots\}$ satisfying $t_c-t_k = \lambda^{-k}\exp(\phi/\omega)$, where $\lambda = \exp(2\pi/\omega) > 1$ is a fundamental scaling ratio and $k$ is an integer. In this way, the LPPLS formula captures the intermittent accelerations and quiescences, with their alternating frequency increasing geometrically as $t_c$ is approached. The existence of log-periodic oscillations is a hallmark of discrete scale invariance[39,47], which represents a partial break of continuous scale invariance, such that the observable remains invariant only when scaling $t_c$–$t$ by a set of specific factors that are integer powers of $\lambda$. Log-periodicity indicates that the system and/or the underlying physical mechanisms are associated with characteristic length/time scales, arising from the localised and threshold nature of the mechanics of rupture in heterogeneous materials[38,43]. It is worth noting that the LPPLS formula includes only the first-order correction to the monotonous power law, since higher-order terms with progressively smaller amplitudes are generally less significant[48].

We implement a two-step parametric calibration scheme to fit the LPPLS equation against the time series data of the observable quantity, briefly summarised as follows (see Appendix A for details). First, we define an optimal time window for the LPPLS fitting by employing the Lagrange regularisation method[49] to detect the starting time $t_0$, with the final time fixed at the last available data point prior to the final event. Second, we estimate the LPPLS model parameters using the ordinary least squares method to minimise the cost function defined as the sum of the squares of the differences between the model and observation[50].

We also perform a nonparametric test to qualify further the existence of log-periodic oscillations in time series data. First, we remove the power law trend in time series by applying the following transformation[51]:

$$\varepsilon(t) = \frac{\Omega(t) - A - B(t_c - t)^m}{C(t_c - t)^m}, \qquad (5)$$

where $\varepsilon$ is the normalised residual that ideally obeys a pure cosine function of the logarithm of the normalised time $\tau = (t_c–t)/(t_c–t_0)$, if the LPPLS formula matches the data perfectly. As the data $\Omega(t)$ is usually sampled regularly in time, the corresponding $\varepsilon$ signals are usually unevenly spaced when using the log time scale $\ln\tau$, making the standard fast Fourier transform unsuitable for our analysis. Thus, we employ the Lomb spectral method[52], which does not require equidistant sampling (see Appendix B), making it well-suited here to detect periodicity on a logarithmic time scale approaching a singularity. From the Lomb periodogram, we can derive a set of parameters to quantitatively characterise log-periodicity and assess its statistical significance, including the dominant angular log frequency $\omega$, the fundamental scaling ratio $\lambda$, the maximum peak height $P_{max}$, the false-alarm probability $p_{FA}$, and the signal-to-noise ratio $\kappa$ (see Appendix B).

## 3. Results

The first case is an underground coal mine located at a depth of ~250 m in Australia[53]. This mine was developed using the long wall mining method. The roof displacement of a gateroad—a rectangular tunnel of 5.2 m wide and 3 m high providing access to the longwall face—was monitored using multipoint extensometers at an accuracy of 0.5 mm. Two catastrophic failure events that occurred on 4th and 11th June 2004 (denoted respectively as event 1 and 2) were recorded at two different locations along the gateroad. A clear precursory accelerating and oscillating pattern in the displacement time series is observed prior to the failure event 1, for which the LPPLS model gives an excellent fit to the data with the intermittency being well captured (Fig. 1a). One can see that this log-periodic behaviour lasted for about 2.5 days before the tunnel roof collapse. Log-periodic oscillations are evident in the $\epsilon$-$\ln\tau$ plot that in general well described by a cosine function (Fig. 1a, inset), while some minor discrepancies might arise from the effects of higher-order harmonics of the oscillations. From the Lomb periodograms (Fig. 1b), we can clearly identify a dominant peak at the angular log frequency $\omega = 4.84$ (with the corresponding scaling ratio $\lambda = 3.67$). Regarding event 2, the LPPLS model also effectively captures the tunnel roof displacement evolution (Fig. 1c), while the presence of log-periodicity is confirmed by the cyclic $\epsilon$-$\ln\tau$ pattern (Fig. 1c, inset). On the Lomb periodogram, a peak is also observed at $\omega = 4.84$ corresponding to $\lambda = 3.67$ (Fig. 1d), which is highly consistent with that for event 1 (Fig. 1b). Interestingly, a harmonic can be found at an angular log frequency of 9.32 (Fig. 1d), which corresponds approximately to the second harmonic $2\omega$, further indicating the importance of log-periodicity in this data.

The second example is a rockburst event that occurred on 4th April 2007 in a deep platinum mine at South Africa[54]. The time-dependent deformational behaviour of the tabular stope has been monitored by mechanical clockwork meters with an accuracy of 0.3 mm. The existence of log-periodicity is demonstrated by the excellent fit of the LPPLS model to the time series of stope closure (Fig. 2a) and the sinusoidal-like signals in the $\epsilon$-$\ln\tau$ plot (Fig. 2a, inset). The Lomb periodogram reveals a dominant peak at $\omega = 4.74$, with the corresponding $\lambda = 3.76$ (Fig. 2b).

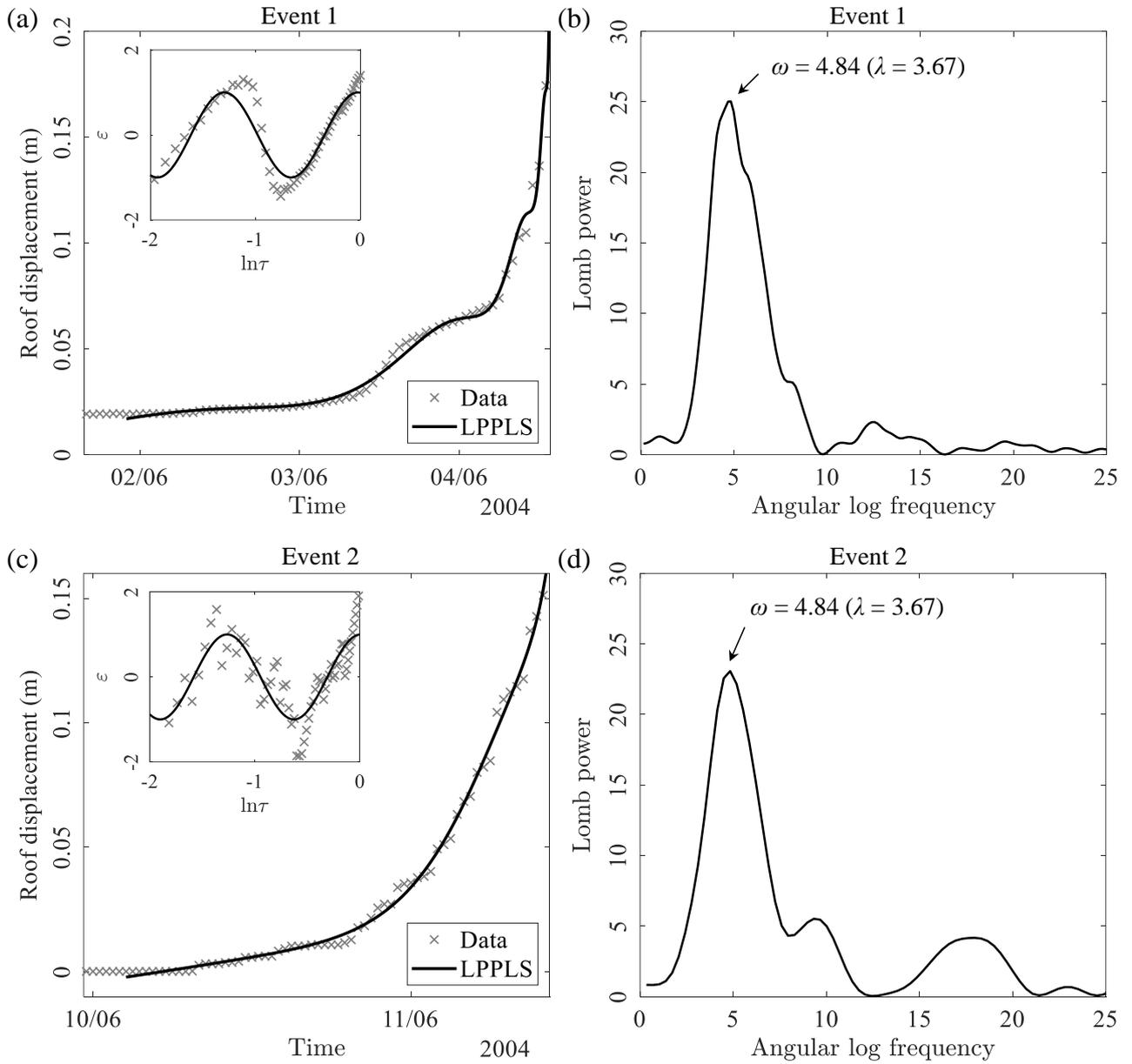

**Fig. 1.** Monitoring data showing the temporal evolution of roof displacement prior to violent failure events on (a) 4th June 2004 (event 1) and (c) 11th June 2004 (event 2) at an underground coal mine in Australia. Insets in the left panel display the variation of normalised residual $\varepsilon$ as a function of the logarithm of normalised time $\tau = (t_c-t)/(t_c-t_0)$. The Lomb periodogram analysis for detecting log-periodic oscillatory components in the data for (b) event 1 and (d) event 2.

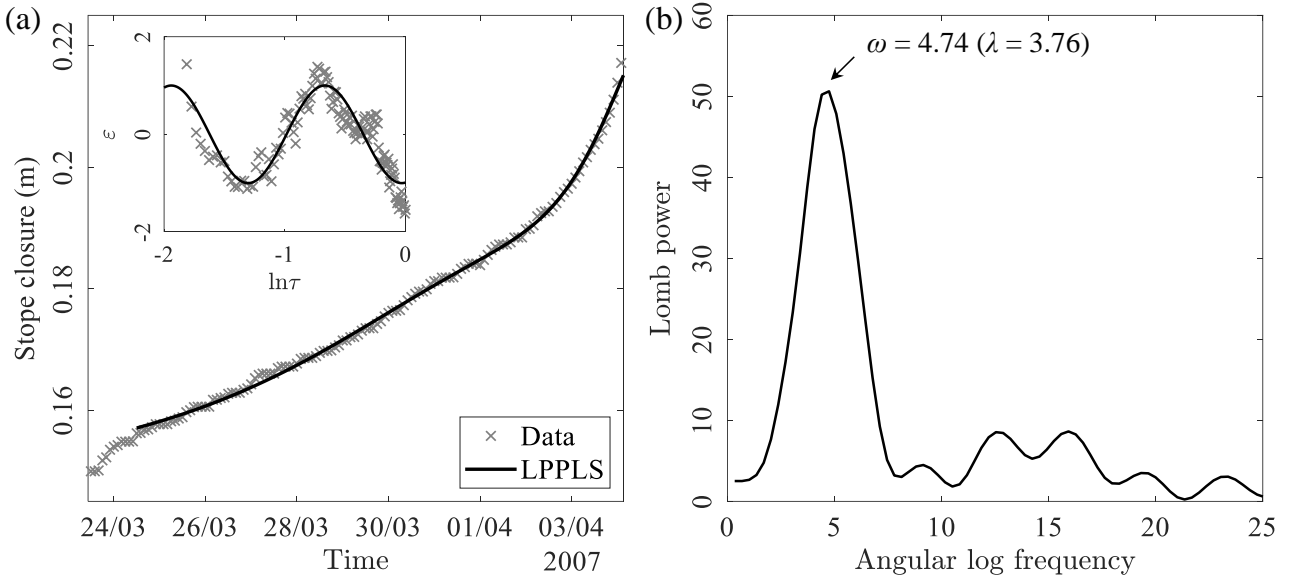

**Fig. 2.** (a) Time series of stope closure measured by a clock meter in a deep platinum mine at South Africa, where a violent rockburst event occurred on 4th April 2007. Inset displays the variation of normalised residual $\varepsilon$ as a function of log normalised time $\tau = (t_c-t)/(t_c-t_0)$. (b) The Lomb periodogram analysis for detecting log-periodic oscillatory components in the data.

The third study case is a deep gold mine at South Africa[40], where eight major rockburst events were documented in 1997. The precursory temporal evolution of Benioff strain release is computed for each of these major events based on the catalogue recorded by a local seismic network[40]. For event 1, the existence of log-periodicity in the Benioff strain data is demonstrated by the LPPLS fit (Fig. 3a, left) and the cyclic $\epsilon$-$\ln\tau$ pattern (inset of Fig. 3a, left), with the Lomb periodogram yielding $\omega =$ 7.13 and $\lambda = 2.41$ (Fig. 3a, right). For event 2, the LPPLS formula also gives a good description of the data (Fig. 3b, left), while the log-peridicity is somewhat still visible in the $\epsilon$-$\ln\tau$ plot (inset of Fig. 3b, left), in spite of the scarcity of the data; the Lomb spectral analysis reveals a dominant peak at $\omega$ = 12.07 with $\lambda = 1.68$ (Fig. 3b, right). For event 3, the significance of log-periodicity is well supported by the excellent LPPLS fit (Fig. 3c, left) and the cyclic $\epsilon$-$\ln\tau$ signals (inset of Fig. 3c, left), while the Lomb periodogram indicates $\omega = 9.42$ with $\lambda = 1.95$ (Fig. 3c, right). For event 4, the LPPLS model in general tracks well the time series data (Fig. 3d, left), with log-periodic structures having significant irregularities in the $\epsilon$-$\ln\tau$ plot (inset of Fig. 3d, left) and a Lomb peak clear observed at $\omega = 9.59$ corresponding to $\lambda = 1.93$ (Fig. 3d, right). For event 5, the precursory log-periodic power law accelerating behaviour is very clear in the data, as reflected by the close match by the LPPLS model (Fig. 3e, left) and the sinusoidal-type signals in the $\epsilon$-ln$\tau$ plot (inset of Fig. 3e, left); the Lomb periodogram reveals a dominant peak at $\omega = 8.81$ with $\lambda = 2.04$ (Fig. 3e, right). For event 6, the existence of log-periodicity is also demonstrated by the LPPLS fit (Fig. 3f, left) and the $\epsilon$-ln$\tau$ pattern (inset of Fig. 3f, left), while the Lomb periodogram reveals one dominant peak at $\omega = 8.42$ and another peak at around $2\omega$ that can be interpreted as the second harmonic, so that the fundamental scaling ratio is $\lambda = 2.11$ (Fig. 3f, right). For event 7, the LPPLS model also provides a good description of the data (Fig. 3g, left), though some scatter is evident, likely due to uncertainties associated with the seismic data, which is also reflected in the $\epsilon$-ln$\tau$ plot (inset of Fig. 3g, right); on the Lomb periodogram, a major peak is found at $\omega = 5.04$ corresponding to $\lambda = 3.48$ (Fig. 3g, right). For event 8, despite of the data scarcity, the log-periodicty is still evident, as demonstrated by the LPPLS fit to the data (Fig. 3h, left) and the cyclic $\epsilon$-ln$\tau$ pattern that closely follows a cosine function (inset of Fig. 3h, left), with a dominant Lomb peak identified at $\omega = 5.74$ indicating $\lambda = 2.99$ (Fig. 3h, right).

In Table 1, we provide a summary of the key parameters derived from the LPPLS parametric calibration and the Lomb nonparametric test for all the 11 rockburst events studied in this work. We can see that the critical exponent *m* in general varies from -1 to 0.6 correspond to $\alpha$ ranging from 1.5 to 3.5. The angular log-periodic frequency $\omega$ falls between 4.5 and 12, with the fundamental scaling ratio $\lambda$ ranging from 1.6 to 3.6 (with six cases around 2 and the other five cases around 3.5), for which the values derived from the LPPLS calibration are in a good consistency with those from the Lomb periodogram. The relative amplitude *C/B* of log-periodic oscillations varies from 0.5% to 15% (with five cases larger than 5%). For the two events at the Australian coal mine as well as the event at the South African platinum mine, the significance of log-periodicity is demonstrated by the large values for the Lomb peak height $P_{max}$, the first-to-second highest peak ratio $\eta$, and the signal-to-noise ratio $\kappa$, in addition to their (near-)zero values for the false-alarm probability $p_{FA}$. For event 1, although the identified $\omega$ is close to the noise-related one $\omega_{noise}$, the high signal-to-noise ratio of 2.27 suggests that the detected log-periodicity is unlikely to originate from noise. For the South African gold mine, among the eight events, three have $P_{max}$ larger than 15, five have $\eta$ greater than 2, and five have $p_{FA}$ not exceeding 0.1, three have $\kappa$ larger than 1.5, and most have the identified $\omega$ far from $\omega_{noise}$, which collectively indicates the presence of genuine log-periodicity.

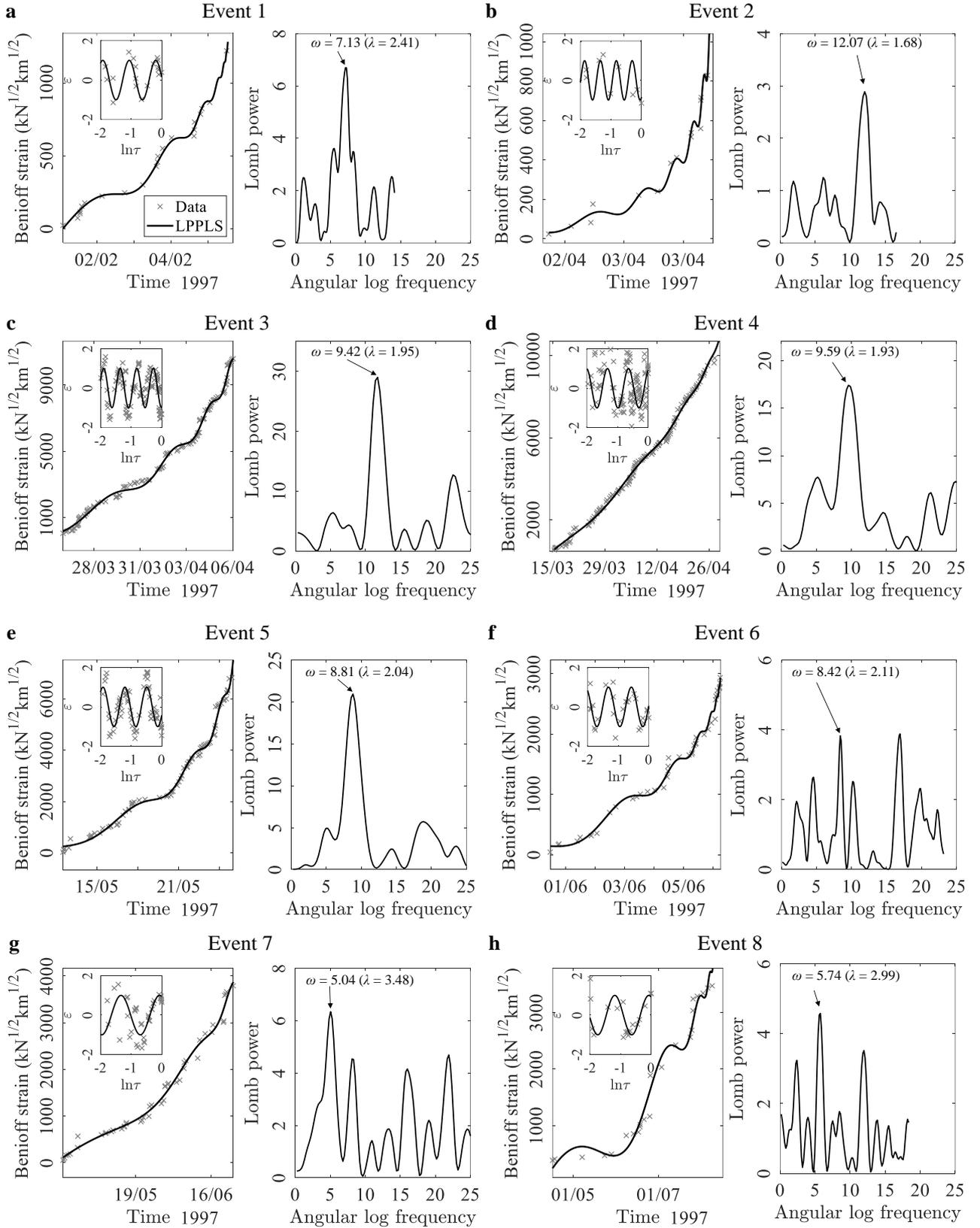

**Fig. 3.** Temporal evolution of Benioff strain release at a deep gold mine at South Africa prior to eight major rockburst events together with the Lomb periodogram analysis for detecting log-periodic oscillatory components in the data. Insets display the variation of normalised residual $\varepsilon$ as a function of logarithm of normalised time $\tau = (t_c-t)/(t_c-t_0)$.

Table 1. Parameters derived from the parametric and nonparametric tests for the 11 rockburst events.

| Event | $m_{LPPLS}$ | $\omega_{LPPLS}$ | $\lambda_{LPPLS}$ | $C/B$ (%) | $\omega_{Lomb}$ | $\lambda_{Lomb}$ | $P_{max}$ | $\eta$ | $p_{FA}$ | $\kappa$ | $\omega_{noise}$ |
|---|---|---|---|---|---|---|---|---|---|---|---|
| Coal mine (event 1) | -0.12 | 4.94 | 3.57 | 2.08 | 4.83 | 3.67 | 25.03 | 10.85 | 0.00 | 2.68 | 2.26 |
| Coal mine (event 2) | -1.1 | 4.94 | 3.57 | 8.17 | 4.84 | 3.67 | 23.07 | 4.18 | 0.00 | 2.27 | 5.15 |
| Platinum mine | -0.08 | 4.94 | 3.57 | 0.44 | 4.74 | 3.76 | 50.62 | 5.86 | 0.00 | 2.75 | 5.09 |
| Gold mine (event 1) | 0.5 | 7.26 | 2.38 | 6.90 | 7.13 | 2.41 | 6.70 | 1.85 | 0.03 | 1.88 | 2.01 |
| Gold mine (event 2) | -0.44 | 12.12 | 1.68 | 6.13 | 12.07 | 1.68 | 2.89 | 2.31 | 0.55 | 1.19 | 3.52 |
| Gold mine (event 3) | 0.07 | 11.61 | 1.72 | 0.48 | 11.71 | 1.71 | 29.02 | 2.28 | 0.00 | 1.52 | 4.78 |
| Gold mine (event 4) | 0.59 | 9.14 | 1.99 | 1.38 | 9.59 | 1.93 | 17.41 | 2.24 | 0.00 | 0.88 | 4.08 |
| Gold mine (event 5) | -0.10 | 8.78 | 2.05 | 0.83 | 8.81 | 2.04 | 20.95 | 3.09 | 0.00 | 1.55 | 3.90 |
| Gold mine (event 6) | 0.42 | 8.37 | 2.12 | 5.22 | 8.40 | 2.11 | 3.83 | 1.01 | 0.20 | 0.70 | 1.88 |
| Gold mine (event 7) | 0.43 | 4.98 | 3.53 | 3.80 | 5.04 | 3.48 | 6.34 | 1.40 | 0.08 | 0.87 | 3.27 |
| Gold mine (event 8) | 0.64 | 5.65 | 3.04 | 15.71 | 5.74 | 2.99 | 4.58 | 1.30 | 0.21 | 1.15 | 2.39 |

Note: $m$ is the critical exponent, $\omega$ is the angular log-periodic frequency, $\lambda$ is the fundamental scaling ratio, $P_{max}$ is the Lomb peak height, $\eta$ is the first-to-second highest peak ratio, $p_{FA}$ is the false-alarm probability, $\kappa$ is the signal-to-noise ratio, and $\omega_{noise}$ is the most probable artificial angular log-periodic frequency. The subscripts "LPPLS" and "Lomb" denote the values obtained from the LPPLS parametric calibration and the Lomb nonparametric test, respectively.

## 4. Discussion

Our analysis of 11 historical rockburst events indicate that the dynamics of observable quantities (displacement, strain, and energy release) characterised by accelerated increase with log-periodic oscillations seem to be a common precursor to violent rockbursts. We have used the log-periodic power law finite-time singularity (LPPLS) model to represent these characteristics and find good agreements between the LPPLS model and the monitoring data across multiple events and mines (Figs. 1-3). Note that, in our previous study, we have carefully ruled out the possibility of overfitting by the LPPLS model as informed by the use of the Akaike and Bayesian information criteria[42]. The significance of log-periodic oscillations is qualified by the sinusoidal shape of residuals (after removing the power law trend) on a logarithmic time scale to the time of the singularity for all the events studied (see insets of Figs. 1-3). In some cases, discrepancies can be observed, which may be either related to the scarcity of data or caused by the presence of higher-order harmonics. These higher-order harmonics are not accounted for in the LPPLS formula that incorporates only the first-order log-periodic correction to the pure power law, but they are identifiable in the Lomb periodograms. The significance of log-periodicity is independently demonstrated by the results of nonparametric Lomb tests. For example, the high Lomb peaks for most rockburst cases, with more than half exceeding 15 and only one marginally below 3 (Table 1), support the statistical significance of log-periodic components[55]. If the data are associated with Gaussian noise, only three events have a false-alarm probability exceeding 0.1 (Table 1). For non-Gaussian noise, the false alarm probability is affected by the distributional and dependence properties of the noise and is in general higher[48], potentially weakening the evidence for genuine log-periodicity. However, the evidence is strengthened by examining the first-to-second Lomb peak ratio (where more than half exceed 2) and by observing consistently high signal-to-noise ratios (mostly above 1), confirming the presence of genuine log-periodicity[48,55]. For most cases, the detected angular log-periodic frequency is very different from the most probable log frequency that can appear from pure noise (Table 1) which, together with the high signal-to-noise ratios, excludes the possible noise origin of the detected log-periodic components.

  We further utilise the detected log-periodic characteristics to gain insigthts into the underlying mechanisms that drive rockburst occurrence. As already mentioned in section 2, log-periodicity is a hallmark of discrete scale invariance[39,47], which is the observable signature of by complex critical exponents generically emerging in nonunitary (dissipative) systems with out-of-equilibrium dynamics and quenched disorder[45]. One possible mechanism for discrete scale invariance is the cascade of ultraviolet Mullins-Sekerka instabilities that could spontaneously emerge in subparallel crack propagation and interaction, where smaller cracks grow more slowly due to stress shadowing effect exerted by nearby fractures, while larger fractures are less affected and grow faster[56]. This stress shadowing effect and fracture interaction phenomenon has been previously reported for an outcrop of a subparallel joint set in a granitic rock at Sierra Nevada[57], while the evidence of log-periodicity has also been discovered in the spacing distribution of an outcrop in granitic rock at Saudi

Arabia[58]. This theory formulated for a system of parallel cracks[56] predicts $\lambda = 2$, which may explain the observed $\lambda \approx 2$ for six rockburst events that occurred in the South African gold mine (Table 1). This aligns with the general view that rockbursts (especially strainbursts) are caused by mode I wing crack propagation, driven by tensile failure along the maximum principal stress direction, subparallel to the excavation boundary[2,12]. This mechanism is expected to dominate in relatively intact rocks, consistent with the competent rock condition in this hard rock mine[40]. Even though rocks inevitably contain some pre-existing flaws, the system may still be approximated as a growing set of subparallel cracks[59,60].

An alternative mechanism (more relevant for fault-slip bursts) to explain $\lambda \approx 2$ is the theory of diffusion-limited aggregation[61,62], which may characterise the diffusive migration of microseismic events in the mine[40], potentially resulting from stress triggering effects[63]. Another potential mechanism involves the interaction between stress drop during seismic phases and stress corrosion during the aseismic periods[64], with a mean-field prediction of $\lambda = 3.6$, which may explain the observed $\lambda \approx 3.5$ for five other rockburst events (see Table 1). This is consistent with another commonly adopted conceptual picture that strainbursts are driven by nonplanar wing crack propagation (driven by stress corrosion) that emanate from the tips of pre-existing fractures experiencing frictional sliding (associated with stress drop) and eventually connect them to form a system-sized rupture plane[65]. For fault-slip bursts, this could correspond to the interplay of stress drop along fault sliding patches and stress corrosion in fault asperities[66]. Some other mechanisms could also contribute to the emergence of log-periodicity, including the interplay of inertia, damage, and healing[67,68] and the interaction of structural heterogeneities and stress concentrations[69]. The pre-existing fracture network in rock may also exhibits a hierarchical pattern[70,71], which may affect log-periodic properties, but discrete scale invariance can spontaneously arise in the absence of any pre-defined hierarchies[47].

Finally, we highlight the potential practical use of log-periodic signals for forecasting rockbursts. By "locking" into the oscillatory rock mass response when approaching catastrophic failure, the LPPLS model can leverage the information of intermittency—traditionally perceived as noise—to constrain and improve the prediction of the timing of rockbursts[42,43]. Moreover, early warning criteria may also be developed for rockburst hazard management based on log-periodic signatures, a method that has been tested in the context of financial bubbles[72].

*Lei & Sornette: Pre-print at arXiv* 13

**Acknowledgement**

D.S. acknowledges partial support from the National Natural Science Foundation of China (Grant No. U2039202, T2350710802), from the Shenzhen Science and Technology Innovation Commission (Grant No. GJHZ20210705141805017) and the Center for Computational Science and Engineering at the Southern University of Science and Technology.

**Appendix A. Parametric LPPLS model calibration**

We implement a stable and robust scheme for calibrating the LPPLS parameters against time series data[49,50]. First. by introducing $C_1 = C\cos\phi$ and $C_2 = C\sin\phi$, we rewrite the LPPLS formula as:

$$\Omega(t) = A + B(t_c - t)^m + C_1(t_c - t)^m \cos[\omega \ln(t_c - t)] + C_2(t_c - t)^m \sin[\omega \ln(t_c - t)], \quad (A1)$$

where the new parameter set $\boldsymbol{\theta}_{\text{LPPLS}} = \{A, B, C_1, C_2, t_c, m, \omega\}$ has seven parameters but with the first four being linear and the last three being nonlinear. This greatly simplifies the calibration by reducing the number of nonlinear parameters from four in the original formula, i.e. equation (4), to three. We then define a cost function as the sum of squared errors:

$$F(\boldsymbol{\theta}_{\text{LPPLS}}; \boldsymbol{\Omega}, \mathbf{t}) = \sum_{i=1}^{N} \left\{\Omega_i - A - B(t_c - t_i)^m - C_1(t_c - t_i)^m \cos[\omega \ln(t_c - t_i)] - C_2(t_c - t_i)^m \sin[\omega \ln(t_c - t_i)]\right\}^2, \quad (A2)$$

where $\boldsymbol{\Omega} = \{\Omega_1, \Omega_2, \ldots, \Omega_N\}$ are $N$ measurements recorded at a series of time stamps $\mathbf{t} = \{t_1, t_2, \ldots, t_N\}$ within a time window $[t_0, t_f]$. We minimise this cost function based on the ordinary least squares method to estimate the LPPLS parameters:

$$\hat{\boldsymbol{\theta}}_{\text{LPPLS}} = \arg\min_{\boldsymbol{\theta}_{\text{LPPLS}}} F(\boldsymbol{\theta}_{\text{LPPLS}}; \boldsymbol{\Omega}, \mathbf{t}). \quad (A3)$$

We enslave the four linear parameters, $A$, $B$, $C_1$, and $C_2$, to the three nonlinear ones, $t_c$, $m$, and $\omega$, to reduce this minimisation problem to:

$$\{\hat{t}_c, \hat{m}, \hat{\omega}\} = \arg\min_{t_c, m, \omega} F_1(t_c, m, \omega), \quad (A4)$$

with

$$F_1(t_c, m, \omega) = \min_{A,B,C_1,C_2} F(A, B, C_1, C_2, t_c, m, \omega) = F(t_c, m, \omega, \hat{A}, \hat{B}, \hat{C}_1, \hat{C}_2). \quad (A5)$$

The estimates for the linear parameters for fixed $t_c$, $m$, and $\omega$ can be obtained from the following minimisation problem for fixed values of the nonlinear ones:

$$\{\hat{A}, \hat{B}, \hat{C}_1, \hat{C}_2\} = \arg\min_{A,B,C_1,C_2} F(A, B, C_1, C_2, t_c, m, \omega), \quad (A6)$$

which can be analytically solved from the following system of equations:

$$\begin{bmatrix} N & \sum f_i & \sum g_i & \sum h_i \\ \sum f_i & \sum f_i^2 & \sum f_i g_i & \sum f_i h_i \\ \sum g_i & \sum f_i g_i & \sum g_i^2 & \sum g_i h_i \\ \sum h_i & \sum f_i h_i & \sum g_i h_i & \sum h_i^2 \end{bmatrix} \begin{bmatrix} \hat{A} \\ \hat{B} \\ \hat{C}_1 \\ \hat{C}_2 \end{bmatrix} = \begin{bmatrix} \sum \Omega_i \\ \sum \Omega_i f_i \\ \sum \Omega_i g_i \\ \sum \Omega_i h_i \end{bmatrix}, \qquad (A7)$$

with $f_i = (t_c - t_i)^m$, $g_i = (t_c - t_i)^m \cos[\omega \ln(t_c - t_i)]$, and $h_i = (t_c - t_i)^m \sin[\omega \ln(t_c - t_i)]$. A filter of $4.94 \leq \omega \leq 15$ is applied during the parametric calibration to exclude physically less relevant oscillation scenarios[46,50].

To determine the optimal time window $[t_0, t_f]$ for the LPPLS calibration, we detect the best start time $t_0$ based on the Lagrange regularisation method[49], with the end time $t_f$ fixed at the last available measurement before the final failure. More specifically, the following cost function is minimised:

$$\tilde{F}'(t_0) = \frac{F}{N(t_0) - 7} - \chi N(t_0), \qquad (A8)$$

where $F$ is given by equation (A2), 7 corresponds to the number of LPPLS parameters, and $\chi$ is the Lagrange multiplier that may be approximated via a linear regression.

There is a possibility that log-periodicity identified from the parametric calibration arises as an artifact due to noise in the data, which can be assessed by examining the calibrated angular log frequency against the most probable artificial one calculated as[73]:

$$\omega_{\text{noise}} = \frac{2\pi \times 1.5}{\ln(t_c - t_0) - \ln(t_c - t_f)}. \qquad (A9)$$

**Appendix B. Nonparametric Lomb spectral analysis**

The Lomb method performs locally least-squares fits to data points of a time series with sinusoids centered at each data point[52]. Given a time series of normalised residual $\varepsilon_i$ sampled at log times $\ln \tau_i$, where $i = 1, 2, \ldots, N$, and $\tau_i = (t_c - t_i)/(t_c - t_0)$, the normalised Lomb periodogram is computed as[55]:

$$P(\omega) = \frac{1}{2\sigma^2} \left\{ \frac{\left[ \sum_1^N (\varepsilon_i - \varepsilon_m) \cos \omega (\ln \tau_i - \xi) \right]^2}{\sum_1^N \cos^2 \omega (\ln \tau_i - \xi)} + \frac{\left[ \sum_1^N (\varepsilon_i - \varepsilon_m) \sin \omega (\ln \tau_i - \xi) \right]^2}{\sum_1^N \sin^2 \omega (\ln \tau_i - \xi)} \right\}, \qquad (B1)$$

with

$$\xi = \frac{1}{2\omega}\arctan\left(\frac{\sum_{1}^{N}\sin 2\omega \ln \tau_i}{\sum_{1}^{N}\cos 2\omega \ln \tau_i}\right), \tag{B2}$$

where $\omega = 2\pi f$ is the angular log frequency, and $\varepsilon_m$ and $\sigma^2$ are respectively the mean and variance of $\varepsilon$. The dominant angular log frequency $\omega$ can be identified in the Lomb periodogram by searching for the maximum peak height $P_{max}$. For data corrupted with independently and normally distributed Gaussian noise, the false-alarm probability $p_{FA}$ for the detected highest peak is given by[55]:

$$p_{FA} = 1 - [1 - \exp(-P_{max})]^M. \tag{B3}$$

where $M$ is the effective number of independent frequencies that can be scanned. The value of $M$ generally depends on the number of sampled frequencies, the number $N$ of data points and the spacing of the data points. However, $M \approx N$ when the data points are approximately equally spaced and the sampled frequencies oversample the available frequency range[74]. Furthermore, the signal-to-noise ratio $\kappa$ can also be computed as[55]:

$$\kappa = \left(\frac{4P_{max}}{N - 2P_{max}}\right)^{1/2}, \tag{B4}$$

which is in general valid for different types of noise[48,55].